\PassOptionsToPackage{unicode}{hyperref}
\PassOptionsToPackage{hyphens}{url}
\PassOptionsToPackage{dvipsnames,svgnames,x11names}{xcolor}
\documentclass[
]{article}
\usepackage{xcolor}
\usepackage[margin=2.5cm]{geometry}
\usepackage{amsmath,amssymb}
\usepackage{iftex}
\ifPDFTeX
  \usepackage[T1]{fontenc}
  \usepackage[utf8]{inputenc}
  \usepackage{textcomp} 
\else 
  \usepackage{unicode-math} 
  \defaultfontfeatures{Scale=MatchLowercase}
  \defaultfontfeatures[\rmfamily]{Ligatures=TeX,Scale=1}
\fi
\usepackage{lmodern}
\ifPDFTeX\else
\fi
\IfFileExists{upquote.sty}{\usepackage{upquote}}{}
\IfFileExists{microtype.sty}{
  \usepackage[]{microtype}
  \UseMicrotypeSet[protrusion]{basicmath} 
}{}
\makeatletter
\@ifundefined{KOMAClassName}{
  \IfFileExists{parskip.sty}{%
    \usepackage{parskip}
  }{
    \setlength{\parindent}{0pt}
    \setlength{\parskip}{6pt plus 2pt minus 1pt}}
}{
  \KOMAoptions{parskip=half}}
\makeatother
\usepackage{longtable,booktabs,array}
\usepackage{calc} 
\usepackage{etoolbox}
\makeatletter
\patchcmd\longtable{\par}{\if@noskipsec\mbox{}\fi\par}{}{}
\makeatother
\IfFileExists{footnotehyper.sty}{\usepackage{footnotehyper}}{\usepackage{footnote}}
\makesavenoteenv{longtable}
\usepackage{graphicx}
\makeatletter
\newsavebox\pandoc@box
\newcommand*\pandocbounded[1]{
  \sbox\pandoc@box{#1}%
  \Gscale@div\@tempa{\textheight}{\dimexpr\ht\pandoc@box+\dp\pandoc@box\relax}%
  \Gscale@div\@tempb{\linewidth}{\wd\pandoc@box}%
  \ifdim\@tempb\p@<\@tempa\p@\let\@tempa\@tempb\fi
  \ifdim\@tempa\p@<\p@\scalebox{\@tempa}{\usebox\pandoc@box}%
  \else\usebox{\pandoc@box}%
  \fi%
}
\def\fps@figure{htbp}
\makeatother
\NewDocumentCommand\citeproctext{}{}

\makeatletter
 \let\@cite@ofmt\@firstofone
 \def\@biblabel#1{}
 \def\@cite#1#2{{#1\if@tempswa , #2\fi}}
\makeatother
\newlength{\cslhangindent}
\newlength{\csllabelwidth}
\newenvironment{CSLReferences}[2] 
 {\begin{list}{}{%
  \setlength{\itemindent}{0pt}
  \setlength{\leftmargin}{0pt}
  \setlength{\parsep}{0pt}
  \ifodd #1
   \setlength{\leftmargin}{\cslhangindent}
   \setlength{\itemindent}{-1\cslhangindent}
  \fi
  \setlength{\itemsep}{#2\baselineskip}}}
 {\end{list}}
\usepackage{calc}

\newcommand{\CSLLeftMargin}[1]{\parbox[t]{\csllabelwidth}{\strut#1\strut}}
\newcommand{\CSLRightInline}[1]{\parbox[t]{\linewidth - \csllabelwidth}{\strut#1\strut}}

\providecommand{\tightlist}{%
  \setlength{\itemsep}{0pt}\setlength{\parskip}{0pt}}
\AtBeginEnvironment{longtable}{\footnotesize\setlength{\tabcolsep}{3pt}}
\usepackage{bookmark}
\IfFileExists{xurl.sty}{\usepackage{xurl}}{} 
\hypersetup{
  colorlinks=true,
  linkcolor={Maroon},
  filecolor={Maroon},
  citecolor={Blue},
  urlcolor={Blue},
  pdfcreator={LaTeX via pandoc}}

\author{}
\date{}

\begin{document}

\section{TiAb Review Plugin: A Browser-Based Tool for AI-Assisted Study
Selection in Systematic
Reviews}\label{tiab-review-plugin-a-browser-based-tool-for-ai-assisted-study-selection-in-systematic-reviews}

\textbf{Authors:}

Yuki Kataoka \textsuperscript{1,2,3,4,5,6}, Masahiro Banno
\textsuperscript{3,7}, Michihito Kyo \textsuperscript{3,8,9}, Shuri
Nakao \textsuperscript{3,10}, Tomoo Sato \textsuperscript{3,11},
Shunsuke Taito \textsuperscript{3,12}, Tomohiro Takayama
\textsuperscript{3,13}, Takahiro Tsuge \textsuperscript{3,14,15},
Yasushi Tsujimoto \textsuperscript{3,16,17,18}, Ryuhei
So\textsuperscript{3,19}, Toshi A. Furukawa \textsuperscript{20}

\textbf{Affiliations:}

\textsuperscript{1} Center for Postgraduate Clinical Training and Career
Development, Nagoya University Hospital, 65, Tsurumai-cho, Showa-ku,
Nagoya-city, Aichi, Japan\\
\textsuperscript{2} Center for Medical Education, Graduate School of
Medicine, Nagoya University, 65, Tsurumai-cho, Showa-ku, Nagoya-city,
Aichi, Japan\\
\textsuperscript{3} Scientific Research WorkS Peer Support Group
(SRWS-PSG), Osaka, Japan\\
\textsuperscript{4} Department of Healthcare Epidemiology, Graduate
School of Medicine, Kyoto University, Yoshida Konoe-cho, Sakyo-ku, Kyoto
606-8501, Japan\\
\textsuperscript{5} Department of International and Community Oral
Health, Tohoku University Graduate School of Dentistry, 4-1,
Seiryo-machi, Aoba-ku, Sendai, Miyagi, 980-8575, Japan\\
\textsuperscript{6} Department of Internal Medicine, Kyoto Min-iren
Asukai Hospital, Tanaka Asukai-cho 89, Sakyo-ku, Kyoto 606-8226, Japan\\
\textsuperscript{7} Department of Psychiatry, Seichiryo Hospital,
Tsurumai 4-16-27, Showa-ku, Nagoya 466-0064, Japan\\
\textsuperscript{8} Department of Radiation Disaster Medicine, Research
Institute for Radiation Biology and Medicine, Hiroshima University,
Hiroshima, Japan\\
\textsuperscript{9} Department of Emergency and Critical Care Medicine,
Graduate School of Biomedical and Health Sciences, Hiroshima University,
Hiroshima, Japan\\
\textsuperscript{10} Division of Rehabilitation Medicine, Shimane
University Hospital, 89-1 Enya-cho, Izumo, Shimane 693-8501, Japan\\
\textsuperscript{11} Department of Nursing, Faculty of Nursing, Kindai
University, 1-14-1 Miharadai, Minami Ward, Sakai City, Osaka, Japan\\
\textsuperscript{12} Division of Rehabilitation, Department of Clinical
Practice and Support, Hiroshima University Hospital, Kasumi 1-2-3,
Minami-ku, Hiroshima, 734-8551, Japan\\
\textsuperscript{13} Kyoto University Hospital, Kyoto, Japan\\
\textsuperscript{14} Department of Rehabilitation, Kurashiki Medical
Center, 250 Bakuro, Kurashiki, Okayama 710-8522, Japan\\
\textsuperscript{15} Department of Epidemiology, Graduate School of
Medicine, Dentistry and Pharmaceutical Sciences, Okayama University,
2-5-1 Shikata-cho, Okayama 700-8558, Japan\\
\textsuperscript{16} Oku Medical Clinic, 7-1-4 Shinmori, Asahi-ku, Osaka
535-0022, Japan\\
\textsuperscript{17} Department of Health Promotion and Human Behavior,
Kyoto University Graduate School of Medicine / School of Public Health,
Kyoto University, Yoshida Konoe-cho, Sakyo-ku, Kyoto 606-8501, Japan\\
\textsuperscript{18} Division of Rheumatology, Department of Internal
Medicine, Showa University School of Medicine, Shinagawa-ku, Tokyo,
Japan\\
\textsuperscript{19} Okayama Psychiatric Medical Center, Okayama,
Japan\\
\textsuperscript{20} Kyoto University Office of Institutional
Advancement and Communications, Kyoto, Japan

\textbf{Corresponding author}\\
\emph{Correspondence}: Yuki Kataoka (youkiti@gmail.com)\\
Address: Center for Postgraduate Clinical Training and Career
Development, Nagoya University Hospital, 65, Tsurumai-cho, Showa-ku,
Nagoya-city, Aichi, Japan

\subsection{Abstract}\label{abstract}

Background: Server-based screening tools impose subscription costs,
while open-source alternatives require coding skills. Full-text
screening, the second stage of study selection, has remained outside the
scope of no-code open-source tools.

Objectives: We developed a browser extension that provides no-code,
serverless artificial intelligence (AI)-assisted study selection,
covering both title and abstract (T\&A) screening and full-text
screening, and examined its functionality.

Methods: TiAb Review Plugin is an open-source Chrome browser extension
(available at
\url{https://chromewebstore.google.com/detail/tiab-review-plugin/alejlnlfflogpnabpbplmnojgoeeabij}).
It uses Google Sheets as a shared database and Google Drive as a PDF
store, requiring no dedicated server and enabling multi-reviewer
collaboration. Users supply their own large language model (LLM) API
key, stored locally and encrypted. For T\&A screening, the tool offers
three modes: manual review, LLM batch screening, and machine learning
(ML) active learning. For full-text screening, it retrieves open-access
PDFs through a cascade of sources (PubMed Central, Europe PMC,
Unpaywall, OpenAlex, and publisher pages), presents each PDF alongside a
decision panel with structured exclusion reasons, supports blinded dual
review with conflict adjudication, optionally submits the PDF to the LLM
for a judgment with page-anchored evidence, and computes PRISMA 2020
flow counts. For ML evaluation, we re-implemented the default ASReview
active learning algorithm (TF-IDF with Naive Bayes) in TypeScript to
enable in-browser execution, and verified equivalence against the
original Python implementation using 10-fold cross-validation on six
datasets. For LLM T\&A evaluation, we compared 16 parameter
configurations across two model families on a benchmark dataset,
validated the optimal configuration (Gemini 3.0 Flash, low thinking
budget, TopP=0.95) with a sensitivity-oriented prompt on five public
datasets (1,038 to 5,628 records, 0.5 to 2.0 percent prevalence), and
subsequently benchmarked nine additional models from four developers on
the tuning dataset.

Results: The TypeScript classifier produced top-100 rankings 100 percent
identical to the original ASReview across all six datasets. For LLM T\&A
screening, recall was 94 to 100 percent with precision of 2 to 15
percent, and Work Saved over Sampling at 95 percent recall (WSS@95)
ranged from 46.3 to 89.3 percent. No additional model exceeded the 96.1
percent recall of the reference configuration on the tuning dataset. The
most recent models tended to trade recall for precision. The full-text
stage is described as implemented. Its classification accuracy has not
yet been evaluated.

Conclusions: We developed a functional browser extension that integrates
LLM screening, ML active learning, and an end-to-end full-text screening
stage into a no-code, serverless environment, ready for practical use in
systematic review study selection.

\subsection{1. Introduction}\label{introduction}

\subsubsection{1.1 The Study Selection Burden in Systematic
Reviews}\label{the-study-selection-burden-in-systematic-reviews}

Systematic reviews employ highly sensitive search strategies designed to
capture all potentially relevant studies, inevitably generating large
volumes of candidate records. Study selection proceeds in two stages:
title and abstract (T\&A) screening, in which each record is evaluated
against predefined eligibility criteria, and full-text screening, in
which the full reports of the remaining records are obtained and
assessed. T\&A screening represents the largest single workload in the
review process and is frequently its primary bottleneck (1,2). An
analysis of 195 systematic reviews registered in PROSPERO found that
searches retrieved between 27 and 92,020 records, with a mean yield of
only 2.94\%, meaning that approximately 97\% of retrieved records were
ultimately excluded (1). Field data from operational reviews report a
median of 2,928 records requiring T\&A screening (range: 651--12,156),
of which 8\% (range: 2--16\%) pass to full-text review, and only 1\%
(range: 0.01--3\%) are included in the final synthesis (2,3). The scale
of this workload has motivated the development of computational tools
that aim to reduce screening effort through relevance prioritization or
automated classification (4,5).

\subsubsection{1.2 Barriers in Existing Screening
Tools}\label{barriers-in-existing-screening-tools}

Despite substantial progress in screening automation, three persistent
barriers limit the accessibility of existing tools: cost, technical
skill requirements, and fragmentation of the workflow across the two
stages of study selection.

\textbf{Cost.} Many screening tools are delivered as cloud-based
software-as-a-service (SaaS) platforms, which must cover the costs of
server infrastructure (authentication, data synchronization, backups,
and security) and ongoing development through subscription fees. A user
survey showed that Covidence and Rayyan ranked highest overall on
features and usability, yet both rely on paid plans for full
functionality (6). Covidence charges up to USD 339 per year for
individual users, depending on the number of reviews. Rayyan's free plan
is limited to three reviews and two reviewers, with premium features
requiring USD 4.99--8.33 per seat per month. DistillerSR starts at USD
19.95 per month for academic users.

\textbf{Technical skill requirements.} The leading open-source
alternative, ASReview, requires Python 3.10 or later, installation via
the command line (\texttt{pip\ install\ asreview}), and a running
command-line interpreter throughout use (7,8). Docker-based deployment
has been introduced to simplify setup, but both Python and Docker remain
unfamiliar to many non-technical researchers, creating a meaningful
adoption barrier.

\textbf{Workflow fragmentation.} Open-source screening tools typically
end at the T\&A stage. ASReview, for example, produces a ranked list of
records but offers no support for obtaining full texts, recording
exclusion reasons, or reconciling full-text decisions between reviewers.
Review teams therefore export their T\&A results and continue in a
separate environment, such as a commercial platform, a reference
manager, or a hand-maintained spreadsheet, in which PDFs are collected
manually and exclusion reasons are tallied by hand for the PRISMA flow
diagram. Each hand-off between tools is a source of transcription error
and lost audit trail.

\subsubsection{1.3 The TiAb Review Plugin}\label{the-tiab-review-plugin}

To address these barriers, we developed the TiAb Review Plugin, a Google
Chrome browser extension that provides LLM-assisted study selection
directly within the researcher's existing web workflow. The tool has
four defining characteristics. First, it operates entirely within the
browser, requiring no software installation, server infrastructure, or
programming skills. Second, it uses Google Sheets as a shared database
and Google Drive as a PDF store, enabling multi-reviewer collaboration
at zero server hosting cost with a built-in audit trail (LLM screening
incurs separate API usage charges; see Section 4.3). Third, for T\&A
screening, it offers three complementary modes (manual review, machine
learning (ML) active learning, and LLM-assisted screening), allowing
researchers to choose the approach that best fits their skills,
resources, and review context. Fourth, it carries the review through the
full-text stage in the same environment, from open-access PDF retrieval,
through blinded dual review with structured exclusion reasons and
optional LLM judgment of the PDF, to PRISMA 2020 flow counts. All API
keys are encrypted and stored locally, and the extension is released as
open-source software under the MIT License.

This version extends version 1 of this preprint, which described the
T\&A stage only.

\subsubsection{1.4 Paper Outline}\label{paper-outline}

The remainder of this paper is organized as follows. Section 2 reviews
related work on screening tools and positions our contribution. Section
3 describes the system architecture, the three T\&A screening modes, and
the full-text screening stage. Section 4 presents the evaluation,
comprising ML equivalence verification, LLM parameter tuning,
cross-dataset LLM validation, and a benchmark of additional LLMs.
Section 5 discusses design choices and future directions.

\subsection{2. Related Work}\label{related-work}

\subsubsection{2.1 Technical Landscape of Screening
Tools}\label{technical-landscape-of-screening-tools}

As of March 2026, the Systematic Review Toolbox listed 49 study
selection tools (9), though some are no longer maintained. These tools
can be categorized along two dimensions: the underlying automation
technique and the deployment model.

Classical ML approaches are exemplified by ASReview (7), which uses
TF-IDF text representation with a Naive Bayes classifier in an active
learning loop and has been shown to reduce screening workload by up to
60\% (10), and Abstrackr (11), which similarly prioritizes records by
predicted relevance. Dense retrieval methods have been explored by
DenseReviewer (4), which applies dense passage retrieval to rank records
by similarity to known relevant studies, outperforming classical active
learning in both effectiveness and efficiency in simulation studies.

LLM-based approaches have emerged rapidly since 2023. A systematic
review and meta-analysis of 15 LLM-based screening models reported
pooled sensitivity of 0.81 (95\% CI: 0.62--0.92) with an area under the
receiver operating characteristic curve (AUROC) of 0.92 (12), while
individual studies have reported sensitivity of 96--100\% with workload
reductions of 40--83\% depending on the model and prompt strategy
(13,14). Most LLM-based screening workflows have nonetheless required
users to write custom scripts for API calls, prompt construction, and
result parsing. Tools such as AISysRev (5) have begun to package these
steps, yet self-hosted deployment still demands programming skills,
limiting accessibility for review teams without software engineering
expertise.

Regarding deployment, server-based SaaS platforms such as Covidence
(15), Rayyan (16), DistillerSR (17), and Elicit (18) provide polished
user interfaces and are widely adopted, but their proprietary backends
limit transparency and reproducibility.

Automation of the full-text stage has received less attention than T\&A
screening, and tool support for it remains limited (19). Full-text
screening requires the report itself, so any automated approach depends
on first obtaining a machine-readable copy, which is only
straightforward for open-access articles. Evaluations of LLMs on
full-text eligibility assessment report a wide range of performance.
Across six thoracic surgery meta-analyses, GPT-4o and Claude 3.5 Sonnet
reached a pooled full-text sensitivity of 0.87 and specificity of 0.96
(20). The GPT-4-based Review Copilot reached full-text sensitivity of
97.6\% with specificity of 47.4\% across four reviews, against 99.2\%
and 83.6\% at the T\&A stage (21). ChatGPT 5.0 reached sensitivity of
0.87 and specificity of 0.70 on 170 full-text articles from one review
(22), and a custom GPT was reliable enough to serve as an assistant
reviewer but not as an autonomous one (19); GPT-4 has also been
evaluated at both stages across languages (14). In these studies the
PDFs were collected by the authors and the model was run through scripts
or a chat interface outside any screening tool; Review Copilot is
released as open-source scripts that take user-supplied PDFs (21).
Open-source tools that combine PDF acquisition, blinded dual review, and
LLM assistance in one place have not, to our knowledge, been described.

\subsubsection{2.2 Unresolved Gaps}\label{unresolved-gaps}

Although the underlying classification methods have advanced
considerably, several gaps remain at the level of tool design and
workflow integration. First, no existing tool combines both no-code
operation and serverless deployment; users must choose between
commercial platforms that abstract away technical complexity (at a
financial cost) and open-source tools that eliminate licensing fees (at
a technical cost). Second, the question of data synchronization and
auditability has been addressed through proprietary backends in SaaS
tools but remains cumbersome in open-source workflows. Third,
open-source tools stop at the T\&A stage, leaving PDF acquisition,
exclusion-reason recording, and full-text conflict resolution to manual
processes outside the tool.

\subsubsection{2.3 Contributions}\label{contributions}

This paper makes four contributions that address the gaps identified
above:

\begin{enumerate}
\def\labelenumi{\arabic{enumi}.}
\tightlist
\item
  \textbf{A browser-embedded study selection tool.} To our knowledge, we
  present the first study selection tool implemented as a browser
  extension, integrating directly into the researcher's existing
  literature search workflow.
\item
  \textbf{Browser-native ML active learning.} We reimplement the
  ASReview default active learning algorithm (TF-IDF + Naive Bayes)
  entirely in TypeScript running in the browser, and verify that it
  produces rankings identical to the original Python implementation.
\item
  \textbf{LLM batch screening validation.} We evaluate the T\&A
  screening performance of the Gemini language model across five
  publicly available datasets with well-defined eligibility criteria,
  showing sensitivity of 94--100\% with 49--88\% of records excluded
  without manual review, and benchmark nine further models from four
  developers under the same protocol.
\item
  \textbf{An end-to-end full-text screening stage.} We describe a
  full-text stage that runs in the same serverless environment: a
  cascade of open-access sources for PDF retrieval, an in-browser PDF
  viewer with structured exclusion reasons, blinded dual review with
  adjudication, LLM judgment of the PDF with page-anchored evidence, and
  automatic PRISMA 2020 flow counts.
\end{enumerate}

\subsection{3. System Description}\label{system-description}

\subsubsection{3.1 User Workflow Overview}\label{user-workflow-overview}

The TiAb Review Plugin is a browser extension for Google Chrome that
integrates directly into the researcher's web-based workflow. Figure 1
illustrates the overall user workflow, which consists of six steps.

\begin{figure}
\centering
\includegraphics[width=1\linewidth,height=0.8\textheight,keepaspectratio,alt={User Workflow Overview.}]{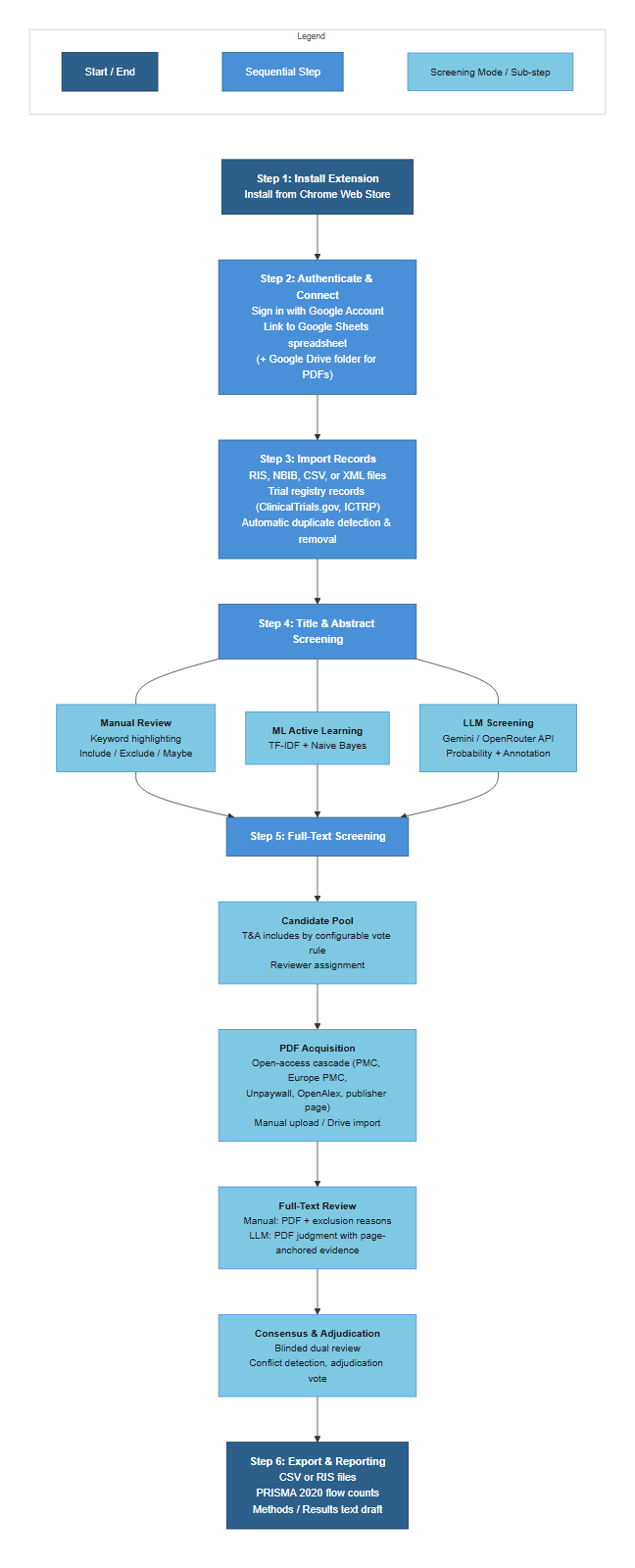}
\caption{User Workflow Overview.}
\end{figure}

First, the user installs the extension from the Chrome Web Store.
Second, the user authenticates with their Google account and connects
the extension to a Google Sheets spreadsheet, which serves as the shared
database for the review project; a Google Drive folder is created
alongside the spreadsheet to hold full-text PDFs. Third, the user
imports bibliographic records by uploading RIS, NBIB, CSV, or XML files
exported from databases such as PubMed or the Cochrane Library, as well
as trial registry exports from ClinicalTrials.gov (CSV) and the WHO
International Clinical Trials Registry Platform (XML); duplicate records
are automatically detected and removed during import. Fourth, the user
conducts T\&A screening using one or more of the three available modes:
manual review, ML-assisted active learning, or LLM screening. These
modes can be used independently or in combination. Fifth, records that
pass T\&A screening enter the full-text stage: the tool retrieves
open-access PDFs, reviewers assess each PDF in an in-browser viewer and
record structured exclusion reasons, the LLM can optionally judge each
PDF, and conflicting decisions are resolved by adjudication. Sixth, the
user exports the results as CSV or RIS files, together with PRISMA 2020
flow counts and a draft of the Methods and Results text for the study
selection section.

\subsubsection{3.2 Architecture}\label{architecture}

The extension is built on Chrome's Manifest V3 platform and comprises
five interconnected components (Figure 2).

\begin{figure}
\centering
\includegraphics[width=1\linewidth,height=0.8\textheight,keepaspectratio,alt={System Architecture. The surrounding frame represents the Chrome Extension boundary (Manifest V3); the dashed frame groups the Google APIs accessed with the user's own account.}]{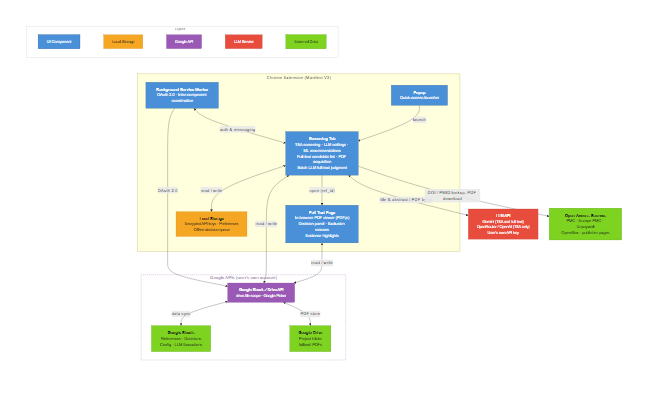}
\caption{System Architecture. The surrounding frame represents the
Chrome Extension boundary (Manifest V3); the dashed frame groups the
Google APIs accessed with the user's own account.}
\end{figure}

The \textbf{background service worker} manages user authentication
through Google's OAuth 2.0 protocol and coordinates communication
between the other components. The \textbf{screening tab} is the primary
user interface where reviewers read titles and abstracts, make screening
decisions, configure LLM settings, and interact with ML recommendations;
it opens as a dedicated tab, providing a full-page workspace. The same
tab hosts the full-text candidate list, PDF acquisition, and batch LLM
judgment of full texts. The \textbf{full-text page} opens as a separate
tab for one record at a time and combines an in-browser PDF viewer
(built on PDF.js) with a decision panel (Section 3.7). The
\textbf{popup} serves as a quick-access launcher for the main interface.
All data exchange with Google Sheets and Google Drive occurs through
Google's Sheets and Drive APIs, using the authenticated user's
credentials under the least-privilege \texttt{drive.file} scope, which
grants access only to files the extension created or the user explicitly
selected through the Google Picker. Local data storage is used for two
purposes: the extension's built-in storage holds encrypted API keys and
user preferences, while a separate browser database (IndexedDB) caches
screening decisions when the network connection is unavailable,
automatically synchronizing them when connectivity is restored.
Open-access PDFs are downloaded by the extension directly from public
sources (Section 3.7.2) and uploaded to the project's Drive folder; the
LLM API receives titles and abstracts (T\&A mode) or PDF bytes
(full-text mode) and nothing else.

\subsubsection{3.3 Data Model}\label{data-model}

The extension uses a Google Sheets spreadsheet as its sole structured
database, organized into four worksheets, and a Google Drive folder as
its file store.

The \textbf{References} sheet stores bibliographic metadata for each
record: a unique identifier (ref\_id), title, abstract, year, authors
(semicolon-delimited, truncated to 10 with ``et al.'' if more), journal,
volume, issue, pages, ISSN, DOI, PubMed ID (PMID), URL, source, import
timestamp, the email address of the importing user, a deduplication key,
the name of the source file from which the record was imported, and a
screening set identifier used for reviewer assignment. The deduplication
key is generated using a priority-based scheme: if a PMID is available,
the key is set to \texttt{pmid:\{pmid\}}; otherwise, if a DOI is
available, the key is \texttt{doi:\{doi\}} (lowercased); failing both, a
normalized version of the title (lowercased, with punctuation and
bracketed annotations removed) is used. The full-text stage adds five
columns: the full-text URL (a Drive link once a PDF has been stored, or
an open-access link when only a link could be found), the full-text
status, the full-text assignment set, and, for PDFs imported from Drive,
the identifiers of the source file and of the copy created in the
project folder. The last two let all team members see that a given PDF
has already been imported. Trial registry records additionally carry a
record type and, once a result publication has been linked to a
registration, a reference to the related record.

The \textbf{Decisions} sheet records every screening decision as a
separate row: a unique decision identifier, the reference identifier,
the reviewer identifier (an email address for human reviewers, a
model-tagged identifier such as \texttt{llm:\{execution\_id\}} for LLM
decisions, or \texttt{adjudication:\{email\}} for a full-text
adjudication vote), the decision itself (include, exclude, maybe, or
pending), a reason field (free text at the T\&A stage; the key of a
structured exclusion reason at the full-text stage), a legacy labels
field retained for backward compatibility, a note field (used to store
structured LLM metadata in JSON format when the decision originates from
automated screening), the timestamp, the client version of the
extension, the URL of the page where the decision was made, and the
screening phase (\texttt{tiab} or \texttt{fulltext}; absent values are
treated as \texttt{tiab} for backward compatibility). This design
supports multiple independent reviewers: each reviewer's decisions are
recorded separately, and conflicts between reviewers are automatically
detected when two or more reviewers have rendered different decisions
for the same reference. A project-level \textbf{Blind} switch controls
whether the extension displays other reviewers' decisions. While it is
on, each reviewer sees only their own decisions and those of the LLM,
and the team progress panel shows decision counts without the
include/exclude breakdown. Because Google Sheets does not natively
enforce access restrictions at the cell level, other reviewers'
decisions remain technically visible in the raw spreadsheet, so
independent screening also relies on an operational protocol in which
reviewers refrain from inspecting the sheet directly until their own
screening is complete. In practice, locating another reviewer's decision
for a specific record within the raw Decisions sheet requires more
effort than independently screening the record.

The \textbf{Config} sheet stores project-level settings as key--value
pairs, including keyword lists for highlighting (with built-in presets
for randomized controlled trials and systematic reviews), LLM screening
parameters (model selection, temperature, thinking level, screening
prompt, inclusion probability threshold, and output language), reviewer
assignment configuration for each stage (calibration set size, group
count, and reviewer-to-reference mapping), the full-text candidate rule,
the project's list of exclusion reasons, the evidence display level for
LLM full-text judgments, the currently adopted LLM full-text round, and
the import statistics (records identified and duplicates removed per
source file) used for the PRISMA counts.

The \textbf{LLM\_Executions} sheet logs each LLM batch operation,
recording the execution identifier, execution type (prompt generation,
T\&A batch screening, or full-text batch screening), timestamp, model
name and parameters (temperature, top-p, thinking level), a snapshot of
the eligibility criteria, the screening prompt used, the inclusion
threshold, counts of targeted, included, and excluded records, a
confirmation status, and an active flag indicating whether the
execution's decisions are currently in use.

The \textbf{Drive folder}
(\texttt{TiAb\ Review\ Plugin/\{project\ name\}/}) contains the
spreadsheet and a \texttt{fulltext/} subfolder holding one PDF per
record. Sharing the folder with a collaborator grants edit access to the
spreadsheet and the PDFs in a single step; an invitation message with
setup instructions can be generated from the share dialog.

\subsubsection{3.4 Manual Review Mode}\label{manual-review-mode}

In manual review mode, the screening tab displays the title and abstract
of the current record, along with three decision buttons: Include,
Exclude, and Maybe. Reviewers navigate between records using on-screen
controls or keyboard shortcuts (I for include, E for exclude, M for
maybe), and each decision is immediately synchronized to the Google
Sheets database. To assist rapid scanning, the extension highlights
user-defined keywords in the displayed text. Two built-in keyword
presets are provided: one for identifying randomized controlled trials
(RCTs) and another for systematic reviews. Users can also define custom
include and exclude keyword lists. The status filter allows reviewers to
view all records, or only those that are pending, included, excluded, or
in conflict between reviewers. When LLM screening has been performed
beforehand, the manual review interface additionally displays the LLM's
inclusion probability and highlights the specific text passages that the
model identified as evidence, enabling the human reviewer to efficiently
verify or override the automated judgment.

\subsubsection{3.5 Machine Learning Active Learning
Mode}\label{machine-learning-active-learning-mode}

The ML mode implements the default active learning algorithm from
ASReview, a widely used open-source tool for systematic review
screening. The original ASReview algorithm uses TF-IDF text
representation combined with a Naive Bayes classifier; we reimplemented
this algorithm entirely in the browser's execution environment to
eliminate the need for any external server or software installation. The
active learning cycle proceeds as follows: (1) the reviewer labels a
small set of seed records (both relevant and irrelevant); (2) the
classifier trains on these labeled examples and predicts the relevance
probability for all remaining records; (3) records are presented in
descending order of predicted relevance, so that the most likely
relevant records are reviewed first; and (4) after each new decision,
the model retrains and the priority order updates. Steps 2 through 4
repeat until a stopping criterion is met. Two stopping criteria are
available: a simple rule that stops after a specified number of
consecutive irrelevant records, and the statistical stopping criterion
of Callaghan and Müller-Hansen(23), which uses a statistical test based
on the hypergeometric distribution to estimate whether a target recall
level (e.g., 99\%) has been reached with a specified confidence (e.g.,
95\%). All model training and prediction runs locally within the browser
using a background computation thread; in ML mode, no bibliographic data
are transmitted to any external service. (In contrast, LLM mode
transmits titles and abstracts to the selected LLM API, and all modes
store data in Google Sheets; see Sections 3.2, 3.3, and 3.6 for
details.)

\subsubsection{3.6 LLM Title and Abstract Screening
Mode}\label{llm-title-and-abstract-screening-mode}

The LLM mode enables automated screening of all records using a large
language model accessed through the user's own API key. Google's Gemini
models are the primary target; models served through OpenRouter (for
example, Qwen and DeepSeek) and OpenAI models can also be selected for
T\&A screening. Two preset Gemini configurations are provided: a
cost-oriented default (Gemini 3.1 Flash Lite at temperature 0; see
Section 4.4) and an accuracy-oriented preset corresponding to the
configuration validated in Section 4.3. The setup begins with the user
pasting their review protocol text, which typically contains structured
eligibility criteria such as population, intervention, comparison, and
outcome (PICO), although any free-text description of the inclusion
criteria can also be used. The extension then uses the LLM to convert
these criteria into an optimized screening prompt, which the user can
review and edit before execution. The screening prompt instructs the
model to evaluate each title--abstract pair against the eligibility
criteria, to assign an inclusion probability between 0 and 1, to provide
a list of reasons for the judgment, and to extract specific text
passages as evidence with their exact character positions. The prompt
explicitly directs the model to favor inclusion when uncertain (``when
in doubt, include''), prioritizing sensitivity over specificity to
minimize the risk of missing relevant studies. The complete prompt
template is available in the source code repository
(\texttt{src/lib/prompt-templates.ts}).

During batch execution, the extension sends each record's title and
abstract to the selected LLM API and stores the full response (inclusion
probability, reasons, and evidence) in the Decisions sheet. Rate
limiting is handled automatically based on the user's API tier. After
the batch is complete, the user sets an inclusion probability threshold
using a slider interface that previews the number of included and
excluded records at each threshold. Because the optimal threshold
depends on the estimated prevalence of relevant records in the dataset,
the interface allows users to adjust the threshold to balance
sensitivity against workload: a lower threshold increases recall at the
cost of more records to manually review, while a higher threshold
reduces workload but risks missing relevant studies. This two-phase
design (automated scoring followed by human threshold confirmation)
ensures that the final screening decisions remain under human control
and can be adapted to the specific risk tolerance of each review.
Furthermore, the stored evidence (inclusion probability, reasons, and
highlighted text passages) is carried forward into the manual review
interface (Section 3.4), where it is displayed alongside the original
title and abstract so that human reviewers can see the basis for the
LLM's judgment when verifying or overriding each decision.

The user's API key is encrypted and stored within Chrome's local
storage. The only data transmitted to the external API are the title and
abstract text of each record; all other processing, including decision
recording and threshold application, occurs locally. Each LLM screening
run is logged in a dedicated LLM\_Executions sheet with its model name,
parameters, prompt version, criteria snapshot, and timestamp, providing
a transparent audit trail that enables full reproducibility of the
automated screening step.

\subsubsection{3.7 Full-Text Screening
Stage}\label{full-text-screening-stage}

The full-text stage is organized as three views within the screening tab
(candidate list, batch LLM judgment, and post-decision review) and a
dedicated full-text page for reading and judging one PDF at a time
(Figure 3).

\begin{figure}
\centering
\includegraphics[width=1\linewidth,height=0.8\textheight,keepaspectratio,alt={The full-text page. The left pane shows the record's bibliographic header, the Drive link of the stored PDF with buttons to replace or delete it, and the PDF rendered in the browser. The right pane holds, from top to bottom, the decision buttons (Include, Exclude, Maybe, with keyboard shortcuts i, e, m), a collapsible panel with the abstract and the reviewer's own T\&A decision, the highlight toggle with the LLM-judgment visibility badge, and the area for the LLM's evidence cards (shown here before any LLM judgment has been run). The footer provides previous, next-undecided, and next navigation. The record and PDF are synthetic demonstration data.}]{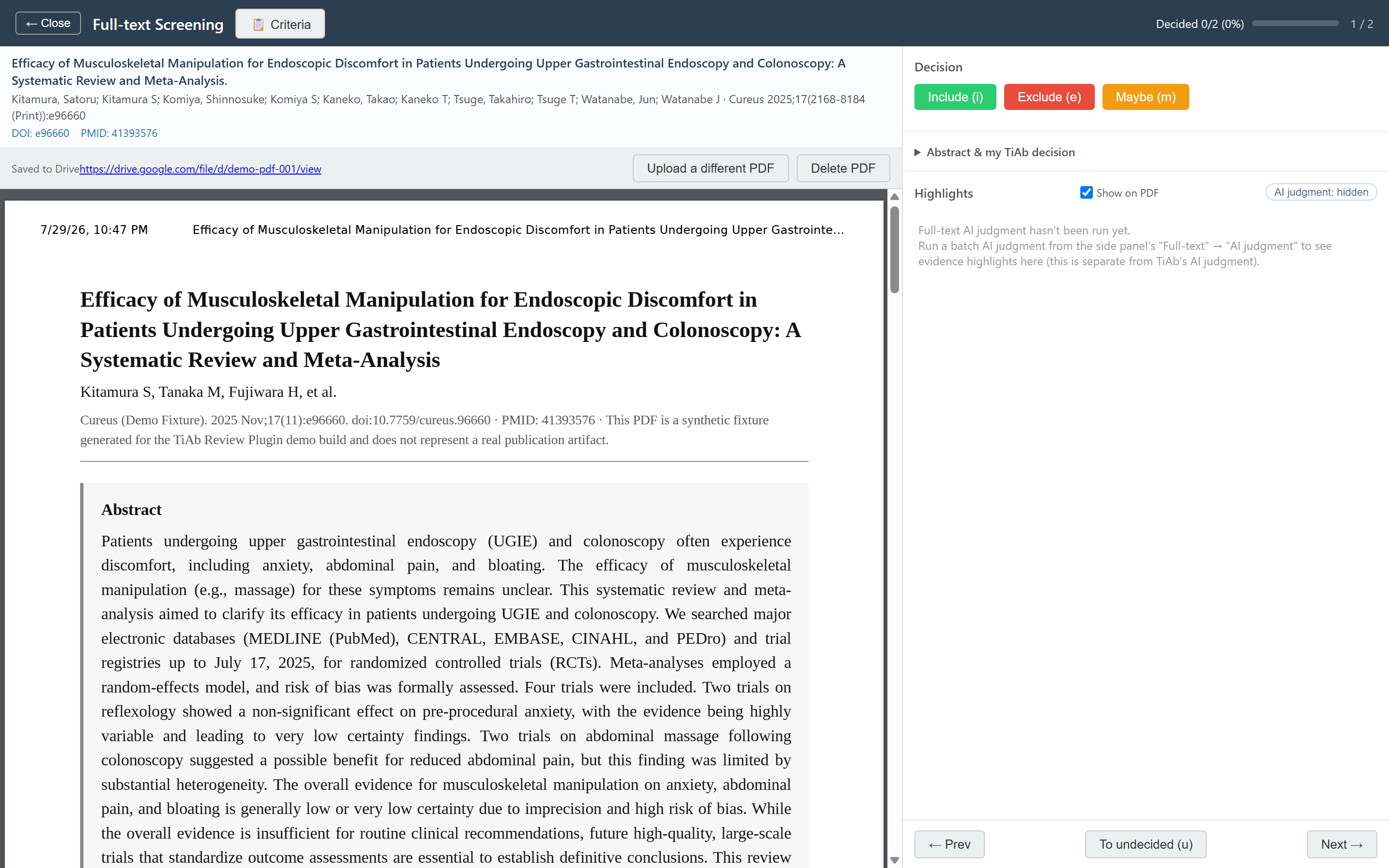}
\caption{The full-text page. The left pane shows the record's
bibliographic header, the Drive link of the stored PDF with buttons to
replace or delete it, and the PDF rendered in the browser. The right
pane holds, from top to bottom, the decision buttons (Include, Exclude,
Maybe, with keyboard shortcuts i, e, m), a collapsible panel with the
abstract and the reviewer's own T\&A decision, the highlight toggle with
the LLM-judgment visibility badge, and the area for the LLM's evidence
cards (shown here before any LLM judgment has been run). The footer
provides previous, next-undecided, and next navigation. The record and
PDF are synthetic demonstration data.}
\end{figure}

\paragraph{3.7.1 Candidate Pool and Reviewer
Assignment}\label{candidate-pool-and-reviewer-assignment}

Records enter the full-text stage according to a project-level candidate
rule. By default, a record becomes a candidate for a reviewer when that
reviewer included it at the T\&A stage (project administrators see every
record that any reviewer included); the rule can instead require a
specified number of Include votes from a chosen set of voters (human
reviewers, the ML model, or particular LLM runs), which lets a team
decide, for example, that a record proceeds when either reviewer
included it. As at the T\&A stage, candidates can be divided into
assignment sets and each set assigned to specific reviewers; records
that enter the pool after the assignment was generated remain visible to
everyone so that none is overlooked. The assignment is stored in the
References sheet rather than derived from decisions, so that all
reviewers see the same candidate list even while decisions are blinded.

\paragraph{3.7.2 PDF Acquisition}\label{pdf-acquisition}

For each candidate, a single button attempts to obtain an open-access
PDF through a cascade of sources, in order: PubMed Central (via the
PMID-to-PMCID converter and the PMC open-access service), Europe PMC,
Unpaywall, OpenAlex, and the \texttt{citation\_pdf\_url} metadata
declared on the publisher's landing page resolved from the DOI. The
first source that yields a downloadable PDF wins; the file is uploaded
to the project's \texttt{fulltext/} Drive folder, and the record's
full-text URL and status are updated. When a source provides a link but
the file cannot be downloaded by the extension (for example, because the
publisher requires a browser session), the record is marked as linked
and the link is shown to the reviewer. Reviewers can also upload PDFs
obtained through institutional subscriptions. Because the
\texttt{drive.file} scope hides files that were not created through the
extension, PDFs placed directly into the Drive folder by a colleague are
imported through the Google Picker. The user selects the files, and the
extension matches each to a record by title, copies it into the project
folder, and records the source and copy identifiers so that the import
is visible to the whole team. The same Picker-based mechanism re-grants
access to PDFs that a team member can no longer read. Before a PDF is
used, the extension checks whether it contains extractable text;
scanned, image-only PDFs are flagged so that reviewers know that
highlight positions may be approximate. For trial registry records, the
extension stores a snapshot of the registration and searches PubMed,
Europe PMC, and the registry's own reference list for result
publications, which can then be imported as linked records.

\paragraph{3.7.3 Manual Full-Text Review}\label{manual-full-text-review}

The full-text page places the PDF viewer on the left and the decision
panel on the right. Reviewers judge with Include, Exclude, or Maybe
using the same keyboard shortcuts as the manual T\&A mode. Choosing
Exclude opens a selector of exclusion reasons; selecting a reason saves
the decision and advances to the next candidate. The reason list is
ordered, and the order is the priority: when several reasons apply, the
reviewer selects the highest-ranked one, which reduces disagreement
between reviewers about the reason. The default list follows PICO
(ineligible population, intervention, comparator, outcome, or study
design; duplicate report; other), and projects can edit the list, for
example to match the PCC framework of a scoping review; each reason
carries an English label used in exports. Choosing Maybe opens a note
field. The panel also shows a collapsible view of the abstract and the
reviewer's own T\&A decision, a ``next undecided'' shortcut, and
progress counters.

\paragraph{3.7.4 LLM Judgment of Full
Texts}\label{llm-judgment-of-full-texts}

From the batch view, the user can submit every cached PDF in the
candidate pool (or a chosen subset) to the Gemini model. Unlike the T\&A
mode, which sends text, the full-text mode sends the PDF bytes
themselves as inline data in the request (up to 18 MB per file); Gemini
reads both text-based and scanned PDFs natively, so no separate OCR step
is required. The model is constrained by a JSON schema to return a
decision (include, exclude, or maybe), an inclusion probability, a
free-text reason, an exclusion-reason category drawn from the project's
own reason list (so that LLM and human reasons are directly comparable
in the PRISMA counts and in conflict detection), and two to five
evidence passages, each with a verbatim quote and the page number on
which it appears. The prompt is generated from the project's eligibility
criteria and the reason list and can be edited before execution. One
batch run constitutes one LLM voter. Its decisions share a reviewer
identifier of the form \texttt{llm:\{model\}@\{timestamp\}}. Each
decision is saved to the Decisions sheet as soon as it is returned, so
that an interrupted run loses none of its completed judgments. A run
that completes without interruption is adopted automatically as the
project's active LLM round. Earlier or interrupted rounds can be adopted
or deleted from a round list. Runs are logged in the LLM\_Executions
sheet with the prompt, model, and counts of judged, failed, and skipped
records.

On the full-text page, the evidence passages are located in the PDF text
layer and highlighted, and each is shown as a card in the decision
panel; clicking a card scrolls the PDF to the passage and vice versa. To
preserve blinding of the human reviewer, the display of LLM output is
governed by a project-level setting with three levels: none (no evidence
shown), neutral (the default: passages are highlighted in a single color
without indicating whether they support inclusion or exclusion, and the
LLM's decision and probability are hidden), and full. Once the Blind
switch is turned off, the decision, probability, and the polarity of
each passage are revealed. Color is supplemented by shape (solid versus
dashed underlines and plus or minus prefixes) so that polarity remains
distinguishable for readers with color vision deficiency.

\paragraph{3.7.5 Consensus, Adjudication, and
Reporting}\label{consensus-adjudication-and-reporting}

The post-decision review view aggregates decisions per reviewer for each
candidate. The consensus rule is permissive: among the selected voters,
any Include yields Include, otherwise any Maybe yields Maybe, otherwise
Exclude. Two kinds of disagreement are flagged separately: conflicting
decisions, and agreement to exclude with differing reasons. Either can
be resolved by an adjudication vote, recorded with the
\texttt{adjudication:} reviewer prefix, which overrides the consensus
and supplies the final reason; when several adjudication votes exist,
the most recent one prevails. From the same view, the user exports CSV
or RIS files, obtains the PRISMA 2020 counts for the identification,
T\&A, and full-text stages (including the number of reports excluded per
reason), and generates a draft of the Methods and Results paragraphs for
the study selection section. In the draft, the numbers, model names, and
parameters are filled in automatically, and the details that the tool
cannot know (such as how disagreements were resolved) are left as
bracketed placeholders.

\subsubsection{3.8 Availability and
Installation}\label{availability-and-installation}

The TiAb Review Plugin is released as open-source software under the MIT
License and is available at
\url{https://github.com/youkiti/tiab-review-plugin}. The extension can
be installed directly from the Chrome Web Store at
\url{https://chromewebstore.google.com/detail/tiab-review-plugin/alejlnlfflogpnabpbplmnojgoeeabij}.
The minimum requirements are Google Chrome (version 90 or later) and a
Google account for authentication and spreadsheet access. For the LLM
screening modes, an API key is additionally required; a Gemini key can
be obtained free of charge from Google AI Studio at
\url{https://aistudio.google.com/}. No programming skills, command-line
tools, or server infrastructure are needed to install or operate the
tool. A companion web application
(\url{https://youkiti.github.io/tiab-review-plugin/app/}) lets
collaborators who cannot install the extension, including users of
tablets, smartphones, and non-Chrome browsers, open a project created
with the extension and take part in T\&A screening; project creation,
record import, ML and LLM screening, and the full-text stage remain
extension-only. A video tutorial showing the tool's usage is available
at \url{https://youtu.be/Yg51rvO9Yrs}. The version described in this
paper is 0.45.0 (September 2026).

\subsection{4. Evaluation}\label{evaluation}

We evaluated the TiAb Review Plugin along three axes: (1) the
equivalence of our TypeScript ML active learning implementation compared
with the original ASReview Python implementation, (2) the classification
performance of LLM batch T\&A screening in terms of sensitivity and
precision, and (3) the performance of alternative LLMs under the same
protocol. The full-text stage has not yet been evaluated for
classification accuracy (Section 4.5).

\subsubsection{4.1 ML Active Learning
Equivalence}\label{ml-active-learning-equivalence}

To verify that our browser-based reimplementation of the ASReview
default algorithm produces results equivalent to the original Python
implementation, we compared the two implementations on six publicly
available benchmark datasets using stratified 10-fold cross-validation.
Folds were generated using stratified sampling to preserve the class
distribution: positive and negative records were each shuffled
independently using a fixed random seed (seed = 42) and distributed
round-robin across 10 folds. In each fold, all records outside the test
set served as labeled training examples, and both implementations
produced a relevance-ranked list for the test set records. We compared
the top-100 records in each ranked list. The fold generation script is
available in the source repository
(\texttt{experiments/asreview/make\_folds.py}).

\textbf{Results.} The TypeScript and Python implementations selected
exactly the same set of top-100 records across all datasets and all
folds (i.e., 100\% overlap in top-100 record IDs). These results
indicate that the browser-based implementation preserves the practical
prioritization behavior of the ASReview-based method for top-ranked
records while operating entirely client-side without requiring
server-side computation or a Python installation.

\subsubsection{4.2 LLM Parameter Tuning}\label{llm-parameter-tuning}

To determine the optimal parameter configuration for LLM batch
screening, we conducted a systematic comparison of 16 conditions across
two model families. We used the depression benchmark dataset (1,993
records; 280 relevant; prevalence 14.1\%) as the tuning set (24). The
primary evaluation metric was the F-beta score with beta = 7
(\(F_{\beta=7}\)), which weights recall approximately 49 times more
heavily than precision. This metric was adopted following a previous
study to evaluate automated abstract screening (25). We developed the
screening prompt using the following template: a fixed template
instructing the model to prioritize sensitivity (``if you are unsure
\ldots{} you MUST include the study''), combined with a brief,
plain-language inclusion criterion derived directly from the dataset
description: ``Include studies on in vivo models of depression (animal
studies). Exclude human studies. Exclude studies where `depression'
refers to respiratory depression, cardiac depression, etc.'' This prompt
and all other screening parameters were designed and finalized before
tuning.

\textbf{Model A: Gemini 2.5 Flash Lite} (8 conditions: temperature
0.0--1.0 crossed with TopP 0.65 and 0.95). Results are summarized in
Table 1.

\textbf{Table 1.} LLM parameter tuning results for Gemini 2.5 Flash Lite
on the depression dataset.

{\def\LTcaptype{none} 
\begin{longtable}[]{@{}
  >{\raggedright\arraybackslash}p{(\linewidth - 12\tabcolsep) * \real{0.1343}}
  >{\raggedright\arraybackslash}p{(\linewidth - 12\tabcolsep) * \real{0.1642}}
  >{\raggedright\arraybackslash}p{(\linewidth - 12\tabcolsep) * \real{0.0597}}
  >{\raggedright\arraybackslash}p{(\linewidth - 12\tabcolsep) * \real{0.1642}}
  >{\raggedright\arraybackslash}p{(\linewidth - 12\tabcolsep) * \real{0.1343}}
  >{\raggedright\arraybackslash}p{(\linewidth - 12\tabcolsep) * \real{0.1194}}
  >{\raggedright\arraybackslash}p{(\linewidth - 12\tabcolsep) * \real{0.2239}}@{}}
\toprule\noalign{}
\begin{minipage}[b]{\linewidth}\raggedright
Condition
\end{minipage} & \begin{minipage}[b]{\linewidth}\raggedright
Temperature
\end{minipage} & \begin{minipage}[b]{\linewidth}\raggedright
TopP
\end{minipage} & \begin{minipage}[b]{\linewidth}\raggedright
Sensitivity
\end{minipage} & \begin{minipage}[b]{\linewidth}\raggedright
Precision
\end{minipage} & \begin{minipage}[b]{\linewidth}\raggedright
\(F_{\beta=7}\)
\end{minipage} & \begin{minipage}[b]{\linewidth}\raggedright
Processing Time
\end{minipage} \\
\midrule\noalign{}
\endhead
\bottomrule\noalign{}
\endlastfoot
A1 & 0.0 & 0.65 & 93.2\% & 47.8\% & 91.5\% & 344s \\
A2 & 0.0 & 0.95 & 93.2\% & 47.8\% & 91.5\% & 332s \\
A3 & 0.3 & 0.65 & 93.2\% & 47.8\% & 91.5\% & 269s \\
A4 & 0.3 & 0.95 & 93.2\% & 47.6\% & 91.5\% & 271s \\
A5 & 0.5 & 0.65 & 93.2\% & 47.8\% & 91.5\% & 342s \\
A6 & 0.5 & 0.95 & 93.2\% & 47.5\% & 91.5\% & 142s \\
A7 & 1.0 & 0.65 & 92.5\% & 48.1\% & 90.8\% & 139s \\
A8 & 1.0 & 0.95 & 92.5\% & 47.3\% & 90.8\% & 135s \\
\end{longtable}
}

Variation across temperature and TopP settings was minimal (sensitivity
difference \textless{} 1 percentage point). Although cost-effective,
this model's sensitivity ceiling of approximately 93\% was judged
insufficient for high-recall screening.

\textbf{Model B: Gemini 3.0 Flash Preview} (8 conditions: thinking level
MINIMAL--HIGH crossed with TopP 0.65 and 0.95). Results are summarized
in Table 2.

\textbf{Table 2.} LLM parameter tuning results for Gemini 3.0 Flash
Preview on the depression dataset.

{\def\LTcaptype{none} 
\begin{longtable}[]{@{}
  >{\raggedright\arraybackslash}p{(\linewidth - 14\tabcolsep) * \real{0.1043}}
  >{\raggedright\arraybackslash}p{(\linewidth - 14\tabcolsep) * \real{0.1217}}
  >{\raggedright\arraybackslash}p{(\linewidth - 14\tabcolsep) * \real{0.1217}}
  >{\raggedright\arraybackslash}p{(\linewidth - 14\tabcolsep) * \real{0.1304}}
  >{\raggedright\arraybackslash}p{(\linewidth - 14\tabcolsep) * \real{0.1304}}
  >{\raggedright\arraybackslash}p{(\linewidth - 14\tabcolsep) * \real{0.1304}}
  >{\raggedright\arraybackslash}p{(\linewidth - 14\tabcolsep) * \real{0.1304}}
  >{\raggedright\arraybackslash}p{(\linewidth - 14\tabcolsep) * \real{0.1304}}@{}}
\toprule\noalign{}
\begin{minipage}[b]{\linewidth}\raggedright
Condition
\end{minipage} & \begin{minipage}[b]{\linewidth}\raggedright
TopP
\end{minipage} & \begin{minipage}[b]{\linewidth}\raggedright
Thinking Level
\end{minipage} & \begin{minipage}[b]{\linewidth}\raggedright
Sensitivity
\end{minipage} & \begin{minipage}[b]{\linewidth}\raggedright
Specificity
\end{minipage} & \begin{minipage}[b]{\linewidth}\raggedright
Precision
\end{minipage} & \begin{minipage}[b]{\linewidth}\raggedright
\(F_{\beta=7}\)
\end{minipage} & \begin{minipage}[b]{\linewidth}\raggedright
WSS@95
\end{minipage} \\
\midrule\noalign{}
\endhead
\bottomrule\noalign{}
\endlastfoot
B1 & 0.65 & MINIMAL & 95.0\% & 85.2\% & 51.2\% & 93.4\% & 70.5\% \\
B2 & 0.95 & MINIMAL & 94.6\% & 86.1\% & 52.6\% & 93.2\% & 69.3\% \\
B3 & 0.65 & LOW & 95.7\% & 85.5\% & 51.9\% & 94.1\% & 71.6\% \\
\textbf{B4} & \textbf{0.95} & \textbf{LOW} & \textbf{96.1\%} &
\textbf{86.3\%} & \textbf{53.4\%} & \textbf{94.6\%} & \textbf{70.8\%} \\
B5 & 0.65 & MEDIUM & 96.1\% & 84.8\% & 50.9\% & 94.4\% & 70.7\% \\
B6 & 0.95 & MEDIUM & 95.0\% & 85.8\% & 52.2\% & 93.5\% & 70.5\% \\
B7 & 0.65 & HIGH & 94.6\% & 85.1\% & 51.0\% & 93.1\% & 68.7\% \\
B8 & 0.95 & HIGH & 95.4\% & 86.1\% & 52.8\% & 93.8\% & 71.0\% \\
\end{longtable}
}

\textbf{Optimal configuration: B4} (TopP = 0.95, thinking level = LOW),
achieving the highest \(F_{\beta=7}\) of 94.6\% with sensitivity of
96.1\%. The thinking model (Model B) outperformed Flash Lite (Model A)
by approximately 3 percentage points in sensitivity and 5 percentage
points in precision. Among thinking levels, LOW provided the best
balance; MEDIUM and HIGH did not yield improvements commensurate with
their increased computational cost. The prompt instruction to ``include
when in doubt'' was likely one reason for the sensitivity above 95\%,
although a formal ablation study comparing performance with and without
this instruction was not conducted.

\subsubsection{4.3 Cross-Dataset LLM
Validation}\label{cross-dataset-llm-validation}

Using the optimal configuration identified in Section 4.2 (condition B4:
Gemini 3.0 Flash Preview, thinking level = LOW, TopP = 0.95, temperature
= 1.0), we evaluated LLM screening performance across five publicly
available benchmark datasets. Each dataset was screened using the
sensitivity-oriented prompt (``when in doubt, include'') with
dataset-specific eligibility criteria. A fixed inclusion probability
threshold of 0.5 was applied to all datasets: records with LLM-assigned
probability \(\geq\) 0.5 were classified as included, and those below
0.5 as excluded.

We used five datasets (CQ1--CQ5) derived from Oami et al. (26), who
compiled labeled datasets from sepsis-related clinical practice
guideline systematic reviews.

\textbf{Results.} Table 3 presents the per-dataset screening
performance. The five datasets ranged from 1,038 to 5,628 records with
prevalence of 0.5--2.0\%. The LLM achieved sensitivity of 94--100\%
across all datasets, with four false negatives out of 16,645 total
records. Precision ranged from 2\% to 15\%, driven primarily by dataset
prevalence: datasets with lower prevalence (higher proportion of
irrelevant records) yielded lower precision, as expected. The proportion
of records excluded by the LLM without manual review (i.e., records
classified as excluded, comprising both true negatives and false
negatives) ranged from 49\% to 88\%. This metric represents the
reduction in the number of records requiring manual screening, not a
measurement of actual time saved. We also computed Work Saved over
Sampling at 95\% recall (WSS@95), which measures the fraction of records
that need not be screened when records are ranked by the LLM-assigned
inclusion probability and screened from highest to lowest until 95\% of
all relevant records have been identified, adjusted for a
random-sampling baseline (7,27). WSS@95 ranged from 46.3\% (CQ3) to
89.3\% (CQ5) across the five datasets.

\textbf{Table 3.} Cross-dataset LLM screening performance (condition B4:
Gemini 3.0 Flash Preview, TopP 0.95, thinking level LOW; inclusion
probability threshold = 0.5).

{\def\LTcaptype{none} 
\begin{longtable}[]{@{}
  >{\raggedright\arraybackslash}p{(\linewidth - 20\tabcolsep) * \real{0.0606}}
  >{\raggedright\arraybackslash}p{(\linewidth - 20\tabcolsep) * \real{0.2222}}
  >{\raggedleft\arraybackslash}p{(\linewidth - 20\tabcolsep) * \real{0.0707}}
  >{\raggedleft\arraybackslash}p{(\linewidth - 20\tabcolsep) * \real{0.0808}}
  >{\raggedleft\arraybackslash}p{(\linewidth - 20\tabcolsep) * \real{0.0909}}
  >{\raggedleft\arraybackslash}p{(\linewidth - 20\tabcolsep) * \real{0.0909}}
  >{\raggedleft\arraybackslash}p{(\linewidth - 20\tabcolsep) * \real{0.0909}}
  >{\raggedleft\arraybackslash}p{(\linewidth - 20\tabcolsep) * \real{0.0808}}
  >{\raggedleft\arraybackslash}p{(\linewidth - 20\tabcolsep) * \real{0.0404}}
  >{\raggedleft\arraybackslash}p{(\linewidth - 20\tabcolsep) * \real{0.0909}}
  >{\raggedleft\arraybackslash}p{(\linewidth - 20\tabcolsep) * \real{0.0808}}@{}}
\toprule\noalign{}
\begin{minipage}[b]{\linewidth}\raggedright
Dataset
\end{minipage} & \begin{minipage}[b]{\linewidth}\raggedright
Topic
\end{minipage} & \begin{minipage}[b]{\linewidth}\raggedleft
Records
\end{minipage} & \begin{minipage}[b]{\linewidth}\raggedleft
Included
\end{minipage} & \begin{minipage}[b]{\linewidth}\raggedleft
Prevalence
\end{minipage} & \begin{minipage}[b]{\linewidth}\raggedleft
Sensitivity
\end{minipage} & \begin{minipage}[b]{\linewidth}\raggedleft
Specificity
\end{minipage} & \begin{minipage}[b]{\linewidth}\raggedleft
Precision
\end{minipage} & \begin{minipage}[b]{\linewidth}\raggedleft
FN
\end{minipage} & \begin{minipage}[b]{\linewidth}\raggedleft
Records excluded (\%)
\end{minipage} & \begin{minipage}[b]{\linewidth}\raggedleft
WSS@95
\end{minipage} \\
\midrule\noalign{}
\endhead
\bottomrule\noalign{}
\endlastfoot
CQ1 & Fluid therapy for sepsis & 5,628 & 113 & 2.0\% & 98.0\% & 50.0\% &
4.0\% & 2 & 49.0\% & 51.3\% \\
CQ2 & Blood pressure targets in sepsis & 3,400 & 17 & 0.5\% & 100.0\% &
70.0\% & 2.0\% & 0 & 69.0\% & 82.1\% \\
CQ3 & Sodium bicarbonate for metabolic acidosis & 1,038 & 16 & 1.5\% &
94.0\% & 58.0\% & 3.0\% & 1 & 57.0\% & 46.3\% \\
CQ4 & Early goal-directed therapy for sepsis & 4,326 & 72 & 1.7\% &
100.0\% & 66.0\% & 5.0\% & 0 & 65.0\% & 78.4\% \\
CQ5 & Restrictive fluid therapy & 2,253 & 41 & 1.8\% & 98.0\% & 90.0\% &
15.0\% & 1 & 88.0\% & 89.3\% \\
\end{longtable}
}

\textbf{API cost.} Token consumption was recorded for every record
during the experiments. At current Gemini Flash API pricing (USD 0.50
per million input tokens and USD 3.00 per million output tokens, with
thinking tokens billed at the output rate), the average cost per record
with the thinking model was approximately USD 0.001, and the total cost
across all five datasets (16,645 records) was approximately USD 23. The
Gemini API also offers a free tier with rate limits that may be
sufficient for smaller reviews.

\subsubsection{4.4 Benchmark of Additional
LLMs}\label{benchmark-of-additional-llms}

Between May and September 2026, as new models were released, we
evaluated nine further models on the depression tuning dataset (1,993
records; 280 relevant) under the protocol of Section 4.2: the same
sensitivity-oriented prompt, the same eligibility criterion, and a fixed
inclusion probability threshold of 0.5. For each model we varied the
sampling or reasoning settings that the model exposes (temperature,
top-p, thinking level, or reasoning effort) and report the condition
with the highest recall; conditions whose projected cost exceeded USD 20
per run were not executed in full. Models were accessed through the
developer's own API (Gemini, GPT) or through OpenRouter (Qwen,
DeepSeek). Results are summarized in Table 4, with the Section 4.2
reference configuration (B4) and the current cost-oriented default shown
for comparison. Cost per 1,000 records is calculated from the recorded
token usage at the list prices in force at the time of each experiment.

\textbf{Table 4.} Best-recall condition of each additional model on the
depression dataset (n = 1,993; 280 relevant; threshold = 0.5).
Specificity is computed from the reported confusion matrix.

{\def\LTcaptype{none} 
\begin{longtable}[]{@{}
  >{\raggedright\arraybackslash}p{(\linewidth - 14\tabcolsep) * \real{0.3000}}
  >{\raggedright\arraybackslash}p{(\linewidth - 14\tabcolsep) * \real{0.1200}}
  >{\raggedright\arraybackslash}p{(\linewidth - 14\tabcolsep) * \real{0.1800}}
  >{\raggedleft\arraybackslash}p{(\linewidth - 14\tabcolsep) * \real{0.0700}}
  >{\raggedleft\arraybackslash}p{(\linewidth - 14\tabcolsep) * \real{0.0800}}
  >{\raggedleft\arraybackslash}p{(\linewidth - 14\tabcolsep) * \real{0.0800}}
  >{\raggedleft\arraybackslash}p{(\linewidth - 14\tabcolsep) * \real{0.0700}}
  >{\raggedleft\arraybackslash}p{(\linewidth - 14\tabcolsep) * \real{0.1000}}@{}}
\toprule\noalign{}
\begin{minipage}[b]{\linewidth}\raggedright
Model (developer)
\end{minipage} & \begin{minipage}[b]{\linewidth}\raggedright
Access
\end{minipage} & \begin{minipage}[b]{\linewidth}\raggedright
Best condition
\end{minipage} & \begin{minipage}[b]{\linewidth}\raggedleft
Recall
\end{minipage} & \begin{minipage}[b]{\linewidth}\raggedleft
Specificity
\end{minipage} & \begin{minipage}[b]{\linewidth}\raggedleft
Precision
\end{minipage} & \begin{minipage}[b]{\linewidth}\raggedleft
\(F_{\beta=7}\)
\end{minipage} & \begin{minipage}[b]{\linewidth}\raggedleft
USD per 1,000 records
\end{minipage} \\
\midrule\noalign{}
\endhead
\bottomrule\noalign{}
\endlastfoot
Gemini 3.0 Flash Preview (Google), reference B4 & Gemini API & Thinking
LOW, TopP 0.95 & 96.1\% & 86.3\% & 53.4\% & 94.6\% &
\textasciitilde1.70 \\
Gemini 3.1 Flash Lite (Google), current default & Gemini API &
Temperature 0, no thinking & 93.6\% & 90.5\% & 61.6\% & 92.6\% &
\textasciitilde0.30 \\
Gemini 3.5 Flash (Google) & Gemini API & Thinking MINIMAL & 93.2\% &
87.3\% & 54.6\% & 91.9\% & 1.93 \\
Gemini 3.6 Flash (Google) & Gemini API & Thinking LOW & 94.6\% & 84.3\%
& 49.6\% & 93.0\% & 1.70 \\
Gemini 3.7 Flash (Google) & Gemini API & Thinking MEDIUM & 91.1\% &
92.2\% & 65.6\% & 90.4\% & 3.27 \\
Gemini 3.8 Flash (Google) & Gemini API & Thinking LOW & 88.9\% & 94.4\%
& 72.2\% & 88.5\% & 0.98 \\
GPT-5.6 (OpenAI) & OpenAI API & Reasoning effort none & 92.9\% & 84.3\%
& 49.1\% & 91.2\% & 1.87 \\
DeepSeek V4 Flash (DeepSeek) & OpenRouter & Default (thinking) & 91.1\%
& 93.0\% & 68.0\% & 90.5\% & 0.38 \\
Qwen3-235B-A22B-2507 (Alibaba) & OpenRouter & Non-thinking & 93.9\% &
83.3\% & 47.9\% & 92.2\% & 0.07 \\
Qwen3.8-27B (Alibaba) & OpenRouter & Non-thinking, temperature 0 &
85.0\% & 95.0\% & 73.7\% & 84.7\% & 0.93 \\
\end{longtable}
}

\textbf{Results.} No model exceeded the 96.1\% recall of the reference
configuration; the closest were Gemini 3.6 Flash (94.6\%), Qwen3-235B
(93.9\%), and Gemini 3.1 Flash Lite (93.6\%). Across the Gemini Flash
line, every generation after 3.0 fell short of the reference recall
(93.2\%, 94.6\%, 91.1\%, and 88.9\% for 3.5 through 3.8), and the two
most recent generations combined the lowest recall with the highest
precision (65.6\% and 72.2\%). The same pattern appeared for GPT-5.6:
raising the reasoning effort from none to medium reduced recall from
92.9\% to 91.4\% and raised precision from 49.1\% to 64.3\%. That is,
the most recent and more deliberative models tended to judge in the
direction of exclusion despite the explicit instruction to include when
in doubt, although with one dataset and one prompt this remains an
observation rather than an established property of these models.
Lowering the threshold recovers recall at the expense of precision (for
Gemini 3.8 Flash, a threshold of 0.05 gives 97.5\% recall at 38.9\%
precision), but at any given recall the newer models were no more
precise than the reference configuration. An OR-combination of the two
cheapest models (Gemini 3.1 Flash Lite and GPT-5.6 with no reasoning)
reached 95.4\% recall at 48.4\% precision, an improvement of 1.8
percentage points in recall over Flash Lite alone at roughly the
combined cost. On the basis of these results, the extension retains
Gemini 3.0 Flash Preview as its accuracy-oriented preset and adopts
Gemini 3.1 Flash Lite, which processes 1,993 records in about 20 seconds
at roughly one sixth of the cost, as the cost-oriented default. Table 4
should be read as a snapshot: model availability and pricing change
frequently, and the experiment scripts in the repository allow the
benchmark to be repeated for new models.

\subsubsection{4.5 Threats to Validity}\label{threats-to-validity}

Several limitations merit consideration when interpreting these results.
Regarding ML evaluation, we verified implementation equivalence only for
the top-100 ranked records; divergence in lower-ranked records, while
unlikely given identical algorithms, was not assessed. Regarding LLM
evaluation, the cross-dataset validation (Section 4.3) used a single
model family (Google Gemini), and the additional models (Section 4.4)
were compared on a single dataset, so their relative ranking on datasets
with lower prevalence is unknown. The benchmark datasets used in this
study are concentrated in critical care medicine with well-defined
eligibility criteria. All evaluations in this study are retrospective,
using datasets with known ground-truth labels. The full-text stage
(Section 3.7) is described as implemented but has not been evaluated: we
have not measured the proportion of candidate records for which an
open-access PDF can be retrieved, nor the sensitivity of the LLM
full-text judgment against reviewer decisions, and such an evaluation
would need to account for the selection of records for which PDFs are
obtainable. No prospective study has yet been conducted to measure the
tool's impact on screening efficiency, reviewer time, or error rates in
a live systematic review project.

\subsection{5. Discussion}\label{discussion}

\subsubsection{5.1 Design Rationale}\label{design-rationale}

The architectural choices underlying the TiAb Review Plugin were driven
by the goal of minimizing barriers to adoption. We chose the browser
extension format because it embeds screening functionality directly into
the environment where researchers already interact with literature
databases. We selected Google Sheets as the data backend because it is
freely available, supports real-time collaborative editing, provides a
revision history that serves as an audit trail, and is already familiar
to many researchers. We adopted the three-mode design (manual, ML, and
LLM) to accommodate the diversity of user skills, institutional
resources, and review contexts: researchers without API keys can use
manual review with keyword highlighting; those who prefer an active
learning workflow can use the ML mode without any external API
dependency; and those with access to an LLM API key can leverage batch
screening for rapid initial classification. These modes can also be
combined; for example, reviewers may use LLM screening for an initial
pass followed by ML-assisted manual review of borderline cases. At the
API prices in Section 4.3, LLM screening of 16,645 records cost about
USD 23, less than one tenth of the annual individual Covidence
subscription cited in Section 1.2.

Extending the tool to the full-text stage required three further
choices. First, we adopted Google Drive as the PDF store, for the same
reasons as Sheets: it is free, shared with the spreadsheet in one step,
and familiar. We deliberately requested only the \texttt{drive.file}
scope, which is classified as non-sensitive by Google and therefore
avoids the security assessment required for broader Drive access, at the
cost of the extension not seeing files that others place in the shared
folder outside the extension. The Picker-based import (Section 3.7.2) is
the workaround for that cost. Second, PDFs are passed to the LLM as-is
rather than converted to text first. This avoids an OCR dependency and
lets the model handle scanned documents, figures, and tables, at the
price of an upload size limit and of highlights that depend on the model
reporting page numbers and quotes accurately. Third, because the LLM's
judgment is available at the moment the human reviewer opens the PDF, we
treated its visibility as a blinding problem: the default display shows
the passages the model attended to without their polarity, so that
reviewers benefit from the navigation aid without being told the answer.
The three display levels make it possible to run a project, or an
experiment, with the model's judgment hidden, neutral, or fully
revealed.

\subsubsection{5.2 Future Work}\label{future-work}

This work suggests several directions for future research and
development. First, a prospective validation study is needed to measure
the tool's impact on screening efficiency, time savings, and error rates
in live systematic review projects. Second, the full-text stage needs
evaluation of its own: the yield of the open-access retrieval cascade,
the sensitivity and precision of the LLM full-text judgment against
reviewer consensus, and the accuracy of the page-anchored evidence,
ideally on a dataset with full-text labels and archived PDFs. Third, the
observation in Section 4.4 that newer models increasingly favor
exclusion suggests that prompt design and threshold selection will need
periodic re-tuning; a standing benchmark run whenever a model is added
would make that visible. Fourth, formal usability testing using
standardized instruments such as the System Usability Scale (SUS) would
provide evidence on the tool's ease of use across different user
populations.

\subsection{6. Availability and
Reproducibility}\label{availability-and-reproducibility}

The TiAb Review Plugin source code, documentation, and all experimental
data are publicly available at
\url{https://github.com/youkiti/tiab-review-plugin} under the MIT
License. The extension can be installed from the Chrome Web Store
without any programming prerequisites. Experimental configurations,
including model parameters, dataset specifications, fold generation
scripts, and the per-model benchmark reports summarized in Table 4, are
provided in the repository's \texttt{experiments/} directory. The T\&A
screening prompt template is located at
\texttt{src/lib/prompt-templates.ts} and the full-text judgment prompt
and output schema in \texttt{src/lib/gemini-fulltext.ts}. Together,
these resources enable full reproducibility of the results reported in
this paper.

\subsection{7. Funding and Conflicts of
Interest}\label{funding-and-conflicts-of-interest}

This work was supported by a Grant-in-Aid for Scientific Research (C)
from the Japan Society for the Promotion of Science (JSPS), grant number
25K13585.

R.S. reports grants from the Osake-no-Kagaku Foundation, speaker's
honoraria from Otsuka Pharmaceutical Co., Ltd., Nippon Shinyaku Co.,
Ltd., and Takeda Pharmaceutical Co., Ltd., outside the submitted work.
T.A.F. has patents 2020-548587 and 2022-082495 pending, and intellectual
properties for Kokoro-app licensed to Mitsubishi-Tanabe. The remaining
authors declare no competing interests.

\subsection*{References}\label{references}
\addcontentsline{toc}{subsection}{References}

\protect\phantomsection\label{refs}

\end{document}